\documentclass[conference]{IEEEtran}
\IEEEoverridecommandlockouts
\usepackage{cite}
\usepackage{amsmath,amssymb,amsfonts}
\usepackage{algorithmic}
\usepackage{graphicx}
\usepackage{textcomp}
\usepackage{xcolor}
\usepackage{url}
 \usepackage{balance}

\graphicspath{{Figures/icorps_national_summer_2025/}}

\begin{document}

\title{Empirical Analysis of Cloud-Edge Infrastructure Complexity: Practitioner Pain Points and Architectural Directions
}

\author{
\IEEEauthorblockN{Pawissanutt Lertpongrujikorn\textsuperscript{1}, Hai Duc Nguyen\textsuperscript{2}, and Mohsen Amini Salehi\textsuperscript{1}}
\IEEEauthorblockA{\textsuperscript{1}High Performance Cloud Computing (HPCC) Lab, University of North Texas (UNT), USA\\
\textsuperscript{2}Argonne National Laboratory and University of Chicago, USA\\
\{pawissanutt.lertpongrujikorn, mohsen.aminisalehi\}@unt.edu, hai.nguyen@anl.gov}
}
\maketitle

\begin{abstract}
The proliferation of cloud, edge, and Internet of Things (IoT) computing has created unprecedented opportunities for distributed applications. However, this architectural shift introduces profound infrastructural complexity, acting as a significant barrier to developer productivity and innovation. In this paper, we present an empirical analysis based on 101 semi-structured interviews across 86 organizations to investigate the state of cloud-native development practices, pain points, and expectations. Our findings quantitatively validate that deployment complexity (38.6\%) and onboarding difficulty (35.6\%) are the dominant operational bottlenecks, while developers heavily prioritize productivity (53.5\%) and automation (44.6\%) over raw performance optimization. Based on these empirical insights, we examine four architectural directions that address the validated pain points: unified object abstractions (Object-as-a-Service), platform engineering via Internal Developer Platforms, declarative AI/ML serving pipelines, and lightweight edge runtimes based on WebAssembly. Furthermore, we detail the multi-stakeholder ecosystem required for adopting novel infrastructure paradigms, emphasizing that security, operational integration, and strict multi-tenant isolation are prerequisites for production readiness. Our results demonstrate that the primary barrier to distributed computing adoption is not execution performance but infrastructural complexity, and that declaratively governed, higher-level abstractions across multiple paradigms offer viable architectural paths toward alleviating it.
\end{abstract}

\begin{IEEEkeywords}
Cloud Computing, Edge Computing, Serverless, Platform Engineering, AI Infrastructure, WebAssembly, Orchestration, Empirical Study, Developer Experience
\end{IEEEkeywords}

\section{Introduction}
\label{sec:intro}

The evolution of computing from centralized cloud environments to highly distributed cloud-edge continuums has expanded the capability of modern applications~\cite{edgecloud2017, gkonis2023survey}. Yet this capability comes at the cost of profound infrastructural complexity: managing microservice architectures~\cite{velepucha2023survey}, orchestrating state and compute, and configuring fragmented cloud services burden developer productivity~\cite{hellerstein2018serverless}, with practitioners managing an average of 14 vendor tools and facing 100-day onboarding cycles~\cite{harness2024dx}.

This burden is amplified by emerging workloads. AI and machine learning systems add model versioning, GPU scheduling, and inference-pipeline management~\cite{kreuzberger2023mlops,paleyes2022challenges}, while platform engineering reflects industry recognition that developer-facing abstractions remain insufficient. These trends suggest that complexity is not confined to one technology but is systemic across modern distributed infrastructure.

While previous work has proposed novel architectures to address individual challenges---such as cold starts in serverless platforms~\cite{golec2024cold,msc} or data replication in edge networks---there remains a critical lack of empirical research examining the holistic pain points experienced by practitioners operating these modern infrastructures. To address this gap, we conducted an extensive customer discovery and empirical validation study involving 101 interviews across 86 organizations, following the NSF I-Corps National program methodologies.

In response to these findings, we examine architectural directions that address validated pain points through higher-level abstractions and declarative governance. This paper makes three contributions: \textbf{C1. Empirical grounding.} We present a rigorous empirical analysis identifying deployment complexity and onboarding difficulty as the primary barriers to distributed infrastructure adoption. \textbf{C2. Expectation gap.} We quantify the gap between practitioner expectations (prioritizing productivity and automation) and the current state of infrastructure orchestration. \textbf{C3. Architectural mapping.} We map the empirical findings to four architectural directions---unified object abstractions (OaaS), platform engineering via Internal Developer Platforms (IDPs), declarative AI/ML serving pipelines, and lightweight edge runtimes (WebAssembly)---and analyze the multi-stakeholder adoption ecosystem, showing that security, operational integration, and multi-tenant isolation are prerequisites for production readiness.

The remainder of this paper is organized as follows. Section~\ref{sec:methodology} outlines methodology and demographics; Section~\ref{sec:findings} reports practitioner challenges and expectations; Section~\ref{sec:directions} maps findings to architectural directions; Section~\ref{sec:discussion} discusses adoption and production readiness; Section~\ref{sec:related} reviews related work; and Section~\ref{sec:conclusion} concludes.

\section{Methodology}
\label{sec:methodology}

\subsection{Customer Discovery Approach}
We adopted the evidence-based customer discovery framework prescribed by the NSF I-Corps program~\cite{blank2020four}, which emphasizes getting out of the building to test hypotheses about customer problems, needs, and workflows. Our approach consisted of semi-structured interviews designed to explore current workflows, pain points, attempted solutions, and desired outcomes rather than pitching our specific technology outright.

\subsection{Interview Protocol}
Each interview followed a consistent protocol: \textbf{Duration:} 30+ minutes per interview; \textbf{Format:} semi-structured conversations via video conference, phone, or in-person; \textbf{Focus areas:} current cloud/edge-cloud development workflows, deployment practices, primary challenges (technical and organizational), performance and cost concerns, prior attempted solutions, and expectations for improvement; and \textbf{Data capture:} participant roles, company industry and size, coded thematic data on specific pain points and desired outcomes, and any quantitative metrics mentioned regarding time savings or performance improvements.

Following each interview, we conducted systematic thematic coding to identify recurring pain points and expectations for improvement. We categorized pain points into eleven primary themes: deployment complexity, onboarding difficulty, system complexity, serverless-specific concerns, security concerns, integration challenges, observability limitations, cost management, scaling issues, documentation gaps, and responsiveness issues. We separately categorized expectations for improvement into thirteen dimensions. This dual coding approach—capturing both current pain points and desired future states—enabled us to validate the problem-solution fit between market needs and edge-cloud orchestration capabilities.

\subsection{Participant Demographics}
Over the course of the study, we conducted 101 interviews that spanned 86 unique organizations (some organizations had multiple interviewees).

\begin{figure}[htbp]
    \centering
    \includegraphics[width=0.48\textwidth]{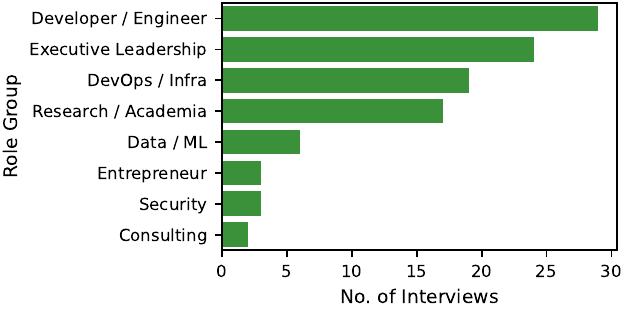}
    \caption{Distribution of interview participants by role group (N=101 interviews). Developer/Engineer and DevOps/Infra roles comprised nearly half of all interviews, reflecting our focus on technical practitioners directly involved in cloud-native development and deployment.}
    \label{fig:role-distribution}
\end{figure}

\textbf{Role Distribution}: As shown in Fig.~\ref{fig:role-distribution}, participants spanned diverse technical and leadership roles. Table~\ref{tab:role-distribution} summarizes the distribution of interview participants by role category. 

\begin{table}[ht]
    \centering
    \caption{Distribution of interview participants by role category (N=101)}
    \small
    \begin{tabular}{|p{2.5cm}|c|p{4cm}|}
        \hline
        \textbf{Role Category} & \textbf{\#Interviews} & \textbf{Specific Roles} \\ \hline \hline
        Dev / Engineer & 29 & Software Eng, Full-stack Devs, Product Eng, QA Eng \\ \hline
        Exec Leadership & 24 & Managers, Founders, CEOs, CTOs, VPs of Eng, PMs \\ \hline
        DevOps / Infra & 19 & DevOps Eng, Infra Eng, SREs, IT Administrators \\ \hline
        Research / Academia & 17 & Professors, Researchers, Research Scientists, PhDs \\ \hline
        Data / ML & 6 & Data Scientists, ML Eng, Research Scientists \\ \hline
        Security & 3 & Security Eng, DevSecOps \\ \hline
        Entrepreneur & 3 & Independent start-up founders \\ \hline
        Consulting & 2 & Technical consultants \\ \hline
    \end{tabular}
    \label{tab:role-distribution}
\end{table}

Industry coverage spanned seven sectors led by Education \& Academia, Software \& Technology, and Financial Services; company sizes ranged from fewer than 10 to over 10,000 employees, ensuring coverage across operational scales.

\section{Empirical Findings: State of the Continuum}
\label{sec:findings}

Our empirical analysis revealed a consistent set of critical, data-supported challenges that quantitatively confirm a significant operability gap within modern cloud-native infrastructures.

\subsection{Pervasive Complexity and Operational Overhead}

\begin{figure}[htbp]
    \centering
    \includegraphics[width=0.48\textwidth]{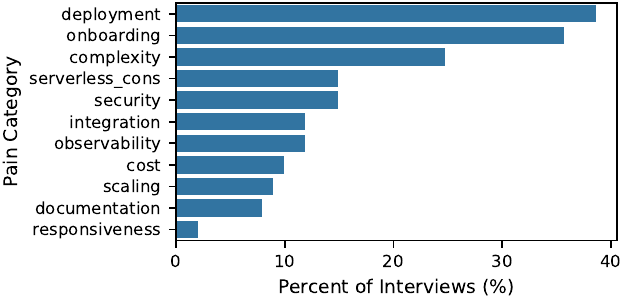}
    \caption{Frequency of pain points mentioned across 101 interviews. Deployment complexity and onboarding difficulty dominated.}
    \label{fig:pain-points}
\end{figure}

Fig.~\ref{fig:pain-points} summarizes the frequency of pain points across all 101 interviews. The top challenges were: \textbf{1) Deployment complexity (38.6\%).} Fragmented tooling, configuration sprawl, and orchestration across compute, storage, and networking forced practitioners to spend significant time on ``glue code'' and infrastructure-as-code templates. \textbf{2) Onboarding difficulty (35.6\%).} Steep learning curves, cloud-specific APIs, and long time-to-productivity created a significant barrier to entry. \textbf{3) System complexity (24.8\%).} The cognitive load of managing microservices dependencies, distributed state, and failure modes overwhelmed many teams. \textbf{4) Security concerns (14.9\%).} Security configuration, credential management, and compliance added overhead, especially in regulated industries with zero-trust requirements~\cite{rose2020zero}. \textbf{5) Serverless-specific concerns (14.9\%).} Cold start latency, execution time limits, and vendor lock-in in Function-as-a-Service platforms~\cite{denninnartefficiency,denninnart2021harnessing} hindered performance guarantees. \textbf{6) Cost management (9.9\%).} Unpredictable costs and opaque cost-performance trade-offs made it difficult to correlate resource usage with business value.

Less frequent concerns included integration challenges (11.9\%), observability limitations (11.9\%), scaling issues (8.9\%), and documentation gaps (7.9\%).

Importantly, these pain points often co-occurred. Our correlation analysis revealed that complexity and onboarding challenges frequently appeared together, identifying the phenomenon of \textit{cognitive overload}---where engineers possess the tools to solve a problem, but orchestrating the tools effectively exceeds reasonable human capital limits.

\subsection{Desired Outcomes and Expectations for Improvement}

\begin{figure}[htbp]
    \centering
    \includegraphics[width=0.48\textwidth]{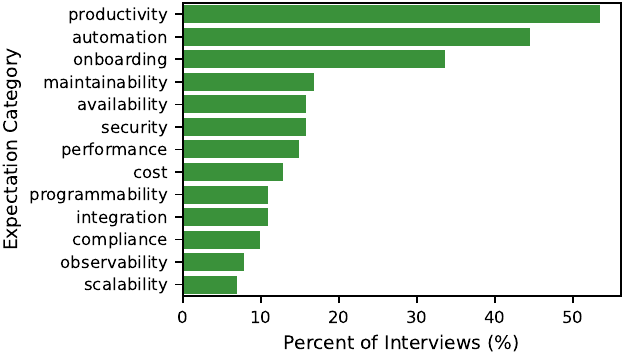}
    \caption{Frequency of expectations for improvement mentioned across 101 interviews. Productivity and automation outrank raw performance metrics.}
    \label{fig:expectations}
\end{figure}

Beyond identifying operational constraints, we codified explicit expectations for improvement across thirteen dimensions (Fig.~\ref{fig:expectations}): \textbf{1) Productivity improvements (53.5\%).} Participants sought quantifiable time-to-value gains across the software delivery lifecycle. \textbf{2) Automation (44.6\%).} They wanted more automation in CI/CD pipelines~\cite{gupta2024continuous}, autoscaling, and rollback processes to remove manual steps and approval bottlenecks. \textbf{3) Onboarding improvements (33.7\%).} Teams expected intuitive abstractions built into developer workspaces to reduce ramp time from months to weeks. \textbf{4) Maintainability (16.8\%).} Respondents wanted safer updates and less organizational toil from brittle automation scripts. \textbf{5) Availability/Reliability (15.8\%) \& Security (15.8\%).} They emphasized reduced MTTR through innate resiliency and simpler data locality enforcement. \textbf{6) Performance (14.9\%).} Lower latency and higher throughput, especially tail latency, mattered but ranked below human-centric usability requirements.

Further expectations included cost optimization (12.9\%), programmability (10.9\%), integration (10.9\%), compliance (9.9\%), observability (7.9\%), and scalability (6.9\%).

Participants were notably less interested in raw hardware-level performance optimization than in \textit{developer experience}. This highlights a notable gap in much contemporary edge-cloud systems research, which often optimizes for nano-second data plane enhancements while underestimating the complexity of the control plane and API abstraction surfaces.

Taken together, these findings point to a common root cause: the fragmentation of abstraction planes forces developers to manually compose independent services for compute, state, and orchestration. Alleviating the dominant pain points---deployment complexity, onboarding difficulty, and cognitive overload---therefore requires higher-level abstractions that consolidate these concerns into declaratively governed deployment units. The following section examines four architectural directions that address this structural deficiency from complementary angles.

\section{Architectural Directions}
\label{sec:directions}

The empirical evidence presented in Section~\ref{sec:findings} identifies a clear demand for abstractions that reduce deployment fragmentation, shorten onboarding cycles, and automate operational toil. No single paradigm is likely to address all practitioner contexts; rather, the findings motivate a family of complementary architectural directions. This section maps the dominant pain points to four such directions: unified object abstractions, platform engineering, AI/ML serving pipelines, and lightweight edge runtimes.

\subsection{Unified Object Abstractions (OaaS)}

\begin{figure}[htbp]
    \centering
    \includegraphics[width=0.48\textwidth]{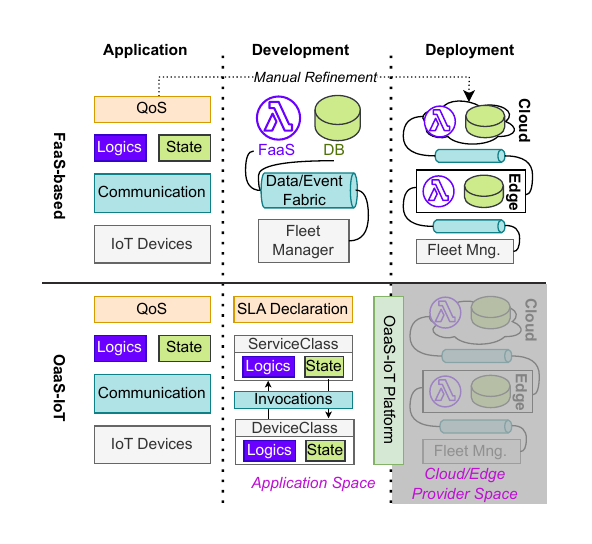}
    \vspace{-5mm}
    \caption{Comparison of FaaS and OaaS paradigms. In FaaS (top), developers must manually integrate external state stores and coordinate communication via data/event fabrics. In OaaS (bottom), compute, state, and workflow are unified into a single object abstraction managed by the platform.}
    \label{fig:faas-vs-oaas}
\end{figure}

Object-as-a-Service (OaaS) consolidates compute, state, and orchestration into a unified object abstraction governed by declarative Non-Functional Requirements (NFRs), directly targeting the fragmented abstraction planes identified as the root cause of practitioner pain, as Figure~\ref{fig:faas-vs-oaas} illustrates. OaaS was initially formulated to abstract data-intensive cloud-native workloads~\cite{lertpongrujikorn2023object}. Subsequent work enriched the model with workflow orchestration and execution guarantees for dispersed environments~\cite{lertpongrujikorn2026object}, including Oparaca-style encapsulation of orchestrator logic~\cite{lertpongrujikorn2026object}, and introduced formal NFR enforcement mechanisms~\cite{lertpongrujikorn2024streamlining}. Most recently, the EdgeWeaver framework extended OaaS to the edge-cloud continuum for IoT scenarios~\cite{lertpongrujikorn2026edgeweaver}.

\subsection{Platform Engineering and Internal Developer Platforms}

The rise of platform engineering as a discipline directly mirrors our empirical findings on developer experience. The concept is rooted in the recognition that excessive cognitive load on development teams is a primary bottleneck for software delivery; a recent practitioner survey found that 93\% of respondents consider platform engineering beneficial for reducing this burden~\cite{puppet2023platform}. Industry surveys report that developers manage an average of 14 distinct vendor tools, contributing to 100-day onboarding cycles~\cite{harness2024dx}---a finding that strongly corroborates our onboarding difficulty statistic. Recent research on developer experience (DevEx) has further established that flow state, feedback loops, and cognitive load are the three key dimensions affecting developer productivity~\cite{greiler2023devex}. Platform engineering addresses these dimensions through \textit{Internal Developer Platforms} (IDPs)~\cite{rusum2024platform}: curated, self-service layers that abstract infrastructure provisioning behind standardized interfaces, as illustrated by developer portals such as Backstage~\cite{backstage2024} and the CNCF survey's evidence of widespread Kubernetes production use~\cite{cncf2024survey}.

\subsection{AI/ML Serving Pipelines}

The deployment challenges identified in our study manifest acutely in AI/ML infrastructure: MLOps introduces additional orchestration complexity from model versioning, GPU scheduling, and inference pipeline management~\cite{kreuzberger2023mlops,paleyes2022challenges}. Although our interview sample included a modest Data/ML cohort (6 of 101 participants), their pain points---cost unpredictability, scaling difficulties, and integration complexity---echo the general findings, with the productivity gap amplified by the distance between experimental notebooks and production serving. Recent work on LLM serving on preemptible GPUs~\cite{miao2024spotserve} illustrates how cost-performance trade-offs mirror the broader demand for declarative, outcome-oriented resource governance. The findings motivate AI serving architectures that integrate model artifacts, data pipelines, and inference endpoints into unified deployment units---paralleling OaaS---with declarative interfaces for latency, throughput, and cost constraints, as emerging platforms like BentoML~\cite{bentoml2024} and KServe~\cite{kserve2024} pursue.

\subsection{Lightweight Edge Runtimes: WebAssembly}

The serverless-specific concerns in our study---cold-start latency, execution time limits, and vendor lock-in---intensify at the edge, where resource-constrained devices cannot support full container runtimes. WebAssembly (Wasm) addresses these directly: near-instant cold starts (microsecond-scale versus hundreds of milliseconds for containers), strong sandboxed isolation, and cross-platform portability from a single binary artifact~\cite{menetrey2022wasi} target two validated pain points---performance unpredictability and deployment fragmentation. Comparative studies confirm competitive throughput with reduced resource footprints on constrained edge hardware~\cite{hall2024wasm}. Wasm is complementary to OaaS and IDP: replacing the container substrate of OaaS prototypes with Wasm runtimes reduces invocation overhead while preserving unified object abstractions, and IDP golden paths can provision Wasm-based edge functions alongside containerized cloud services for a consistent developer experience across the continuum.

Across these directions, the original design space also exposes three cross-cutting research gaps. First, higher-level abstractions must avoid hiding performance-critical controls when expert operators need them, suggesting adaptive interfaces that default to declarative policies but expose lower-level tuning selectively. Second, cross-tier workflow guarantees remain difficult when edge nodes are intermittently connected and workloads span stateful objects, ML pipelines, and Wasm functions. Third, ecosystem maturity depends on standardized NFR governance interfaces, idiomatic SDKs, integrated debugging, and observability instrumentation; without these operational surfaces, promising runtimes risk remaining research prototypes rather than deployable platform components.

\section{Discussion}
\label{sec:discussion}

The architectural directions presented in Section~\ref{sec:directions} address practitioner pain points at the technical level. However, real-world adoption requires simultaneously satisfying three internal personas: \textit{Financial Decision Makers} evaluating ROI, \textit{Technical Decision Makers} assessing architectural safety, and \textit{End Users} evaluating ergonomic benefits. Misalignment among these stakeholders was a recurring theme, with developers favoring cognitive-load reduction while technical leadership demanded production-grade maturity.

Our respondents identified three non-negotiable production prerequisites for any of the four directions: (i) \textit{security}, with multi-tenant isolation and audited IAM controls required in regulated industries (financial services, healthcare) before any consolidation architecture is considered; (ii) \textit{idiomatic developer ergonomics}, with familiar SDKs for Python, Java, Go, and TypeScript and IDE integration required to address onboarding difficulty; and (iii) \textit{observability integration}, with distributed tracing and metrics compatible with existing stacks (Prometheus, OpenTelemetry) as a prerequisite for operations teams. Research prototypes targeting any of the four directions face limited commercial adoption without these operational interfaces---a gap that platform engineering is uniquely positioned to bridge.

We identified two distinct adopter profiles. \textit{SMEs and startups} face acute ``infrastructure tax'' without dedicated DevOps teams and are natural early adopters of unified abstractions---OaaS or turnkey ML serving platforms---where one deployment unit replaces manual service composition. \textit{Enterprise adopters} require mature IAM, multi-tenant isolation, and audit logging; for them, IDP-based adoption pathways---delivering novel abstractions through an established internal platform with governance controls---are more viable than direct adoption of research prototypes. Across both profiles, respondents favored open, non-proprietary platforms and cloud marketplace integration models that preserve architectural control while reducing configuration overhead.

Practitioners expect \textit{consumption-based} economic models where cost correlates directly to declared quality of service. When a platform accepts a declarative SLA---whether an OaaS NFR, a KServe autoscaling policy, or an IDP resource quota---resource scaling must behave predictably. Admission control mechanisms that dynamically adjust resources without over-provisioning address the demand for transparent cost-performance mapping across all organizational sizes.

\section{Related Work}
\label{sec:related}

Our work intersects several active research threads.

Several surveys have examined the practical difficulties of modern distributed systems. Soldani et al.~\cite{soldani2018pains} conducted a systematic grey literature review identifying common pains and best practices in microservice architectures, reporting integration testing and fault diagnosis as dominant concerns—findings our interview data corroborates and extends to deployment-level complexity. Luo et al.~\cite{luo2021characterizing} characterized serverless application patterns and failures, revealing that misconfiguration and tooling fragmentation are leading causes of production incidents. Eismann et al.~\cite{eismann2021state} surveyed serverless adoption and found that vendor lock-in and debugging difficulty rank among the top barriers—consistent with our finding that serverless-specific concerns affected 14.9\% of respondents. Our study distinguishes itself by spanning the full cloud-edge continuum rather than focusing on a single platform paradigm, and by systematically coding both pain points and expectations for improvement across 101 practitioner interviews.

Efforts to simplify distributed application development have produced several abstraction models. Function-as-a-Service platforms such as AWS Lambda~\cite{jonas2019cloud} offload infrastructure management but impose statelessness, forcing developers to externalize state management. Actor-based frameworks, including Microsoft Orleans~\cite{bykov2011orleans} and Akka, encapsulate state and behavior into virtual actors with location-transparent messaging. While actors address state co-location, they do not provide declarative quality-of-service enforcement or built-in workflow orchestration. CNCF standards like Dapr~\cite{dapr2024} offer portable building blocks (state stores, pub/sub, bindings) but still require explicit service composition. OaaS differs from these paradigms by unifying compute, state, and workflow into a single deployable object with declarative NFR governance, directly targeting the fragmented abstraction planes our empirical findings identified as the root cause of deployment complexity.

Orchestrating workloads across heterogeneous edge-cloud tiers has been addressed by several frameworks. KubeEdge~\cite{xiong2018extend} extends Kubernetes to edge nodes but retains its imperative, container-centric configuration model. OpenFaaS and OpenWhisk bring serverless semantics to edge environments, yet inherit the statelessness limitations of traditional FaaS. Osmotic computing~\cite{villari2016osmotic} introduced the concept of automatic workload migration between cloud and edge based on resource availability, but lacks formal QoS enforcement mechanisms. WebAssembly has emerged as a lightweight alternative execution substrate; M\'{e}n\'{e}trey et al.~\cite{menetrey2022wasi} demonstrated Wasm as a common layer for the cloud-edge continuum, achieving microsecond cold starts with strong isolation guarantees. Our work complements these systems by demonstrating, through empirical evidence, that practitioners demand declarative, outcome-oriented abstractions rather than additional imperative configuration surfaces.

The operational challenges of deploying machine learning models in production have been extensively studied. Kreuzberger et al.~\cite{kreuzberger2023mlops} provide a comprehensive MLOps taxonomy, while Paleyes et al.~\cite{paleyes2022challenges} systematically catalog deployment lifecycle challenges, reporting that infrastructure concerns often dominate over model accuracy issues. The growing complexity of LLM serving infrastructure~\cite{miao2024spotserve} has further intensified these challenges. While these works examine ML-specific deployment pain, our study provides cross-domain empirical validation demonstrating that the same root causes---fragmented abstractions and manual orchestration---persist across the full cloud-edge continuum, not only in ML pipelines.

Platform engineering has emerged as a discipline focused on reducing developer cognitive load through Internal Developer Platforms~\cite{rusum2024platform}. Industry reports document the scale of this problem: developers managing numerous vendor tools with lengthy onboarding cycles~\cite{harness2024dx}, and organizations reporting that dominant challenges have shifted from technical to cultural and cognitive dimensions~\cite{cncf2024survey}. While Section~\ref{sec:directions} discusses how IDPs address the specific pain points from our study, the related work here positions our empirical contribution: unlike industry surveys that focus on cloud-native early adopters, our interview-based approach captures practitioner perspectives across the full spectrum of distributed infrastructure, including edge and IoT contexts where platform engineering remains nascent.

\section{Conclusion}
\label{sec:conclusion}
This paper presented an empirical investigation rooted in structured interviews across 86 technology organizations, characterizing the operational challenges that define the modern cloud-edge continuum. Our findings indicate that the primary barrier to advanced distributed computing is not execution latency, but rather the infrastructure complexity that significantly impedes the practitioners responsible for deployment.

By quantifying that deployment complexity (38.6\%) and onboarding difficulty (35.6\%) are the dominant pain points---while productivity (53.5\%) and automation (44.6\%) are the most desired improvements---we have established an empirical foundation for evaluating architectural responses. We examined four complementary directions: Object-as-a-Service (OaaS), which unifies compute, state, and workflow into declaratively governed objects; platform engineering via Internal Developer Platforms, which provides self-service abstraction layers; declarative AI/ML serving pipelines, which address the amplified complexity of GPU-bound workloads; and WebAssembly-based edge runtimes, which eliminate cold-start penalties and deployment fragmentation across heterogeneous tiers.

Our analysis of the multi-stakeholder adoption ecosystem further reveals that security, observability integration, and economic transparency are prerequisites for production deployment---requirements that apply across all four directions. Going forward, systems research should prioritize developer experience alongside raw performance, pursue standardized declarative governance interfaces that span paradigms, and bridge the gap between academic prototypes and production-ready platforms.

\section*{ACKNOWLEDGEMENT}
This project is supported by NSF through CNS CAREER Award\# 2419588 and NSF i-Corps Award\# 2524083.

\balance
\bibliographystyle{IEEEtran}
\bibliography{refs}

\end{document}